\documentclass[a4paper, 10pt]{article}

\usepackage[utf8]{inputenc}
\usepackage{authblk}
\usepackage[english]{babel}
\usepackage{times}
\usepackage{hyperref}
\usepackage{graphicx}
\usepackage{parskip}
\usepackage{color}
\usepackage{enumitem}
\usepackage{a4wide}
\usepackage{authblk}

\usepackage[font=small,labelfont=bf]{caption}

\title{The Westermo test system performance data set}

\author{Per Erik Strandberg}
\author{Yosh Marklund}
\affil{Westermo Network Technologies AB, Västerås, Sweden}
\date{November 2023}

\begin{document}

\maketitle

\section*{Abstract}
There is a growing body of knowledge in the computer science, software engineering, software testing and software test automation disciplines. However, a challenge for researchers is to evaluate their research findings, ideas and tools due to lack of realistic data. This paper presents the Westermo test system performance data set. More than twenty performance metrics such as CPU and memory usage sampled twice per minute for a month on nineteen test systems driving nightly testing of cyber-physical systems has been anonymized and released. The industrial motivation is to spur work on anomaly detection in seasonal data such that one may increase trust in nightly testing. One could ask: If the test system is in an abnormal state -- can we trust the test results? How could one automate the detection of abnormal states? The data set has previously been used by students and in hackathons. By releasing it we hope to simplify experiments on anomaly detection based on rules, thresholds, statistics, machine learning or artificial intelligence, perhaps while incorporating seasonality. We also hope that the data set could lead to findings in sustainable software engineering.

\textbf{Keywords:} software engineering, anomaly detection, open data, nightly testing, data visualization, seasonal data, embedded systems, cyber-physical systems, sustainable software engineering.

\section{Specifications}

\begin{description}    
\item[Subject:]{Computer Science -- Embedded Systems, and Software Engineering}
\item[Specific subject area:]{Performance metrics from servers in test systems.}
\item[Type and format of data:]{Tabular data in 19 CSV files, each containing 23 or 24 time series sampled about 86 thousand times. Total size is about 360 MB.}
\item[Data collection process:]{Performance data were acquired from servers using node exporter, stored with grafana and then exported to CSV.}
\item[Data accessibility:]{Available at GitHub:
\href{https://github.com/westermo/test-system-performance-dataset}{https://github.com/westermo/test-system-performance-dataset}
}
\item[Related research article:]{
P E Strandberg. (2021).
\emph{Automated System-Level Software Testing of Industrial Networked Embedded Systems,}
PhD Thesis, Mälardalen University.
ISBN: 978-91-7485-529-6. \cite{strandberg2021automated}
}
\end{description}

\newpage
\section{Introduction}
\begin{figure}
  \begin{center}
    \includegraphics[width=0.8\linewidth]{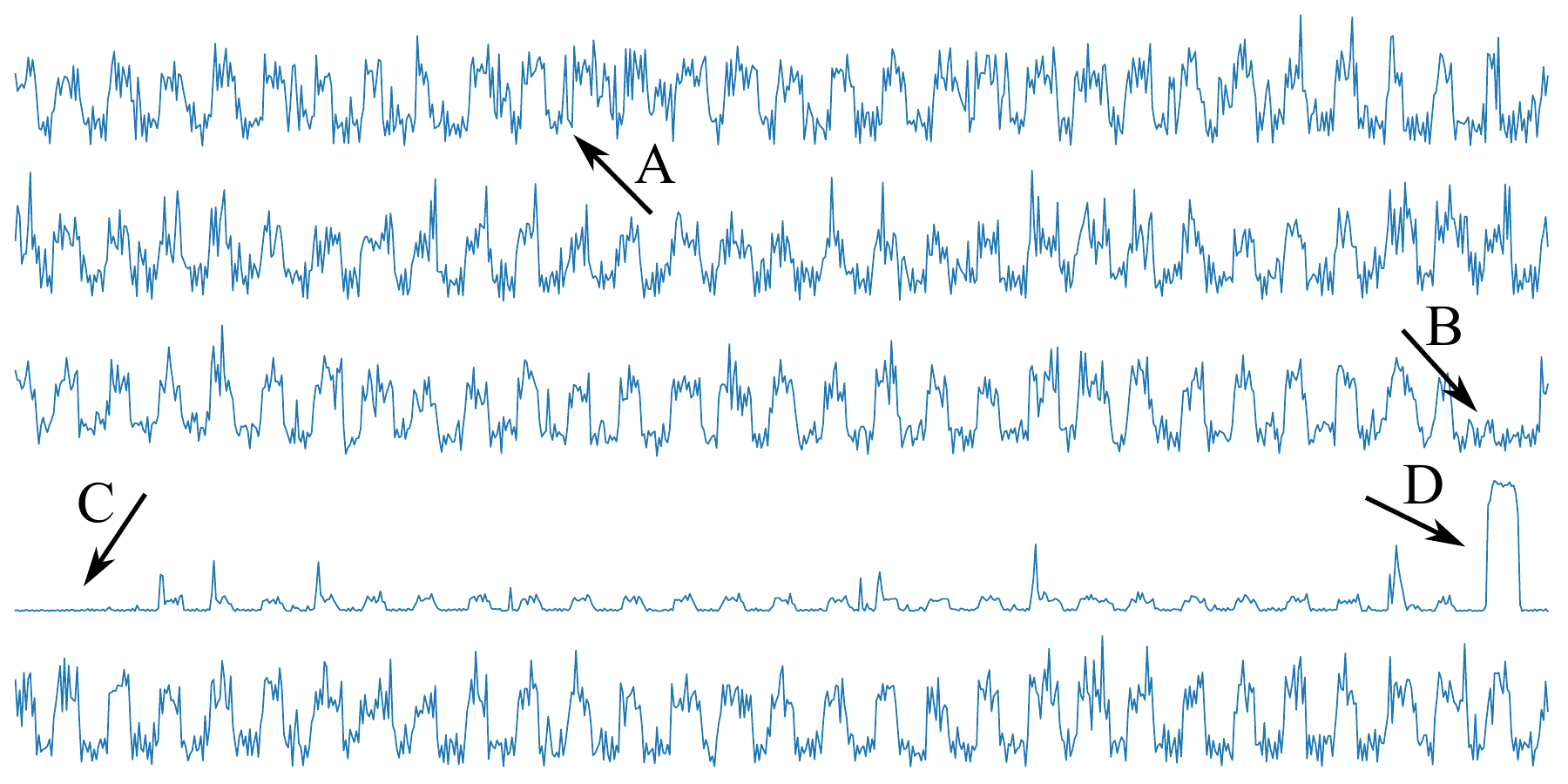}
    \caption{The metric load-15m on five systems (averaged for smoother curves).}
    \label{fig:load15}
  \end{center}  
\end{figure}

There is a growing body of knowledge in the computer science, software engineering, software testing and software test automation disciplines. However, there is a challenge for researchers to evaluate their research findings, innovations and tools due to lack of realistic data. This paper presents the Westermo test system performance data set. This is a data set with performance metrics from test systems that drive nightly testing at Westermo. In Figure~\ref{fig:load15} one such metric is visualized from five test systems. As can be seen, the signal has obvious peaks during the nights, when the test systems are working, and periods of inactivity during the days when humans might use the test systems for debugging and development. 

\begin{figure}
  \begin{center}
    \includegraphics[width=0.8\linewidth]{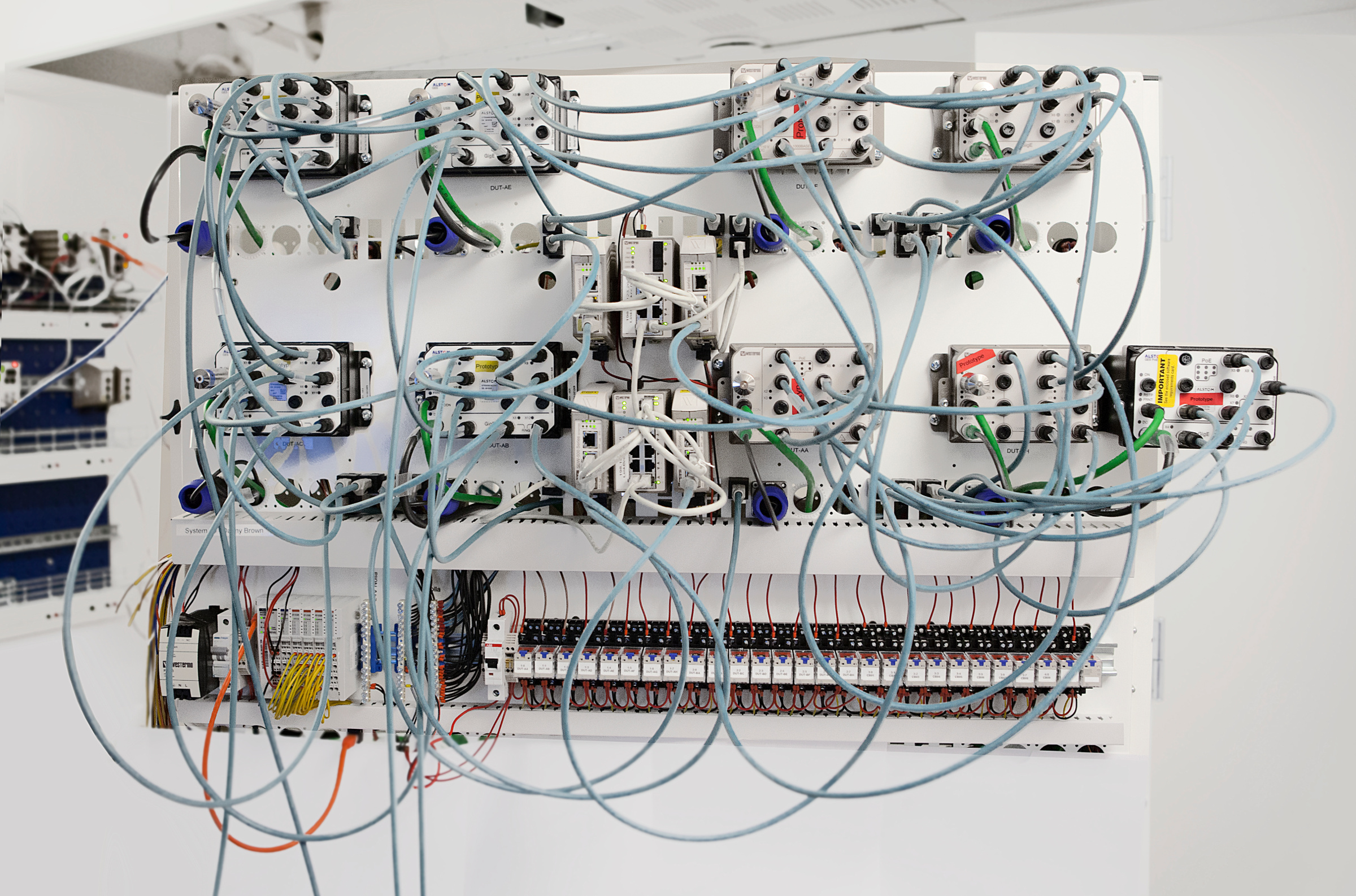}
    \caption{A Westermo test system with a network topology built up of switches, routers, other peripheral equipment, as well as a server (not shown).}
  \label{fig:test-system}
  \end{center}
\end{figure}

Westermo develops cyber-physical systems for industrial communication such as robust switches and routers. These can be used in energy distribution, on-board rail, track-side rail, or for industry automation. A typical product supports many network communication protocols. In addition to robust hardware, the devices run an operating system based on GNU and Linux with the addition of proprietary code and third party code. This software is developed and maintained by several teams of software developers, and to ensure it has quality, Westermo has invested heavily on software test automation. One way of testing is automated functional testing where a test framework configures a test system, enables various protocols, sends certain stimuli and observes the effect such that it may give a pass or fail verdict -- thereby answering "does this software seem to work as expected?" See Figure~\ref{fig:test-system} for an example of one such test system. The test framework runs on a server running Ubuntu GNU/Linux (not shown in the figure).

If the test framework does not give the verdict pass after a test case, this could have many reasons. By design, the desired cause is that the software under test has a problem of some kind. However, with a large and complex test framework, it could be the case that the problem is instead located in the code of the test framework. Furthermore, the root cause could be located in the hardware setup, e.g., if a cable has been connected in the wrong port, or a peripheral device was left without power, etc. Finally, it could be the cause that the problem comes from the server driving the nightly testing -- a trivial cause could be a full disk. In order to learn more about automated detection of such anomalies we are releasing this data set.

The release of this data set follows the release of two previous data sets: the Westermo test results data set \cite{strandberg2022westermo} and the Westermo network traffic data set \cite{strandberg2023westermo}. Both these data sets have led to progress in knowledge in the form of validated ideas, algorithms and tools.

The Westermo test system performance data set was created during the AIDO\-aRt project \cite{eramo2021aidoart} to explore anomaly detection. Figure~\ref{fig:load15} illustrates the load-15m performance metric over time for five test systems. Potential anomalies that a system could warn for include: (A) Did nightly testing not stop (the resting time of the test system is small)? (B) Did nightly testing not start (there seems to be a skipped heartbeat)? (C) Was this system under construction or not in use for three days (the curve looks flat)? (D) Is this a sign of abnormal load (a peak much higher than normal peaks)?

The dataset has been used in AIDOaRt hackathons as well as by a summer intern with the goal of replicating work done in a master's thesis \cite{salahshour2022software} where isolation forest was one method used to detect outliers. Currently it is being used by a group of students in a project course in software engineering. These students will experiment with simulating test systems. First, the simulated systems will replay real data, but some additional test systems will be simulated to create abnormal data in order to trigger anomaly detection. The data produced will be stored in a database, such that a human operator could visualize various trends. We anticipate that visualizations of time series from many parallel metrics on many parallel test systems is non-trivial (in particular since tools like MS Excel rapidly crashes for even relatively trivial tasks and visualizations when exploring data from just one test system). Finally, the students will develop an anomaly detection toolkit could be developed to run in batches or in real time. 

In addition, the data set could be used to work on sustainable software engineering \cite{becker2014karlskrona}. One could ask: what is the carbon footprint of running nightly testing? How much waste do we get in terms of energy overuse, test results that cannot be trusted or human effort lost to debugging? Perhaps this data set could be used to shed light on some of these topics?

Users of the data could explore research questions on how to best define or combine anomaly detection based on rules, thresholds, statistics, machine learning or artificial intelligence, perhaps while incorporating seasonality\footnote{\href{https://en.wikipedia.org/wiki/Seasonality}{https://en.wikipedia.org/wiki/Seasonality}}. In short, students, researchers and practitioners in the field of software engineering, statistics, software test automation, graphic design or artificial intelligence can benefit from the data. It can be used to evaluate algorithms, tools or visualizations for improved validity and generalizability.

\section{Data description}
\begin{figure}
  \begin{center}
    \includegraphics[width=0.8\linewidth]{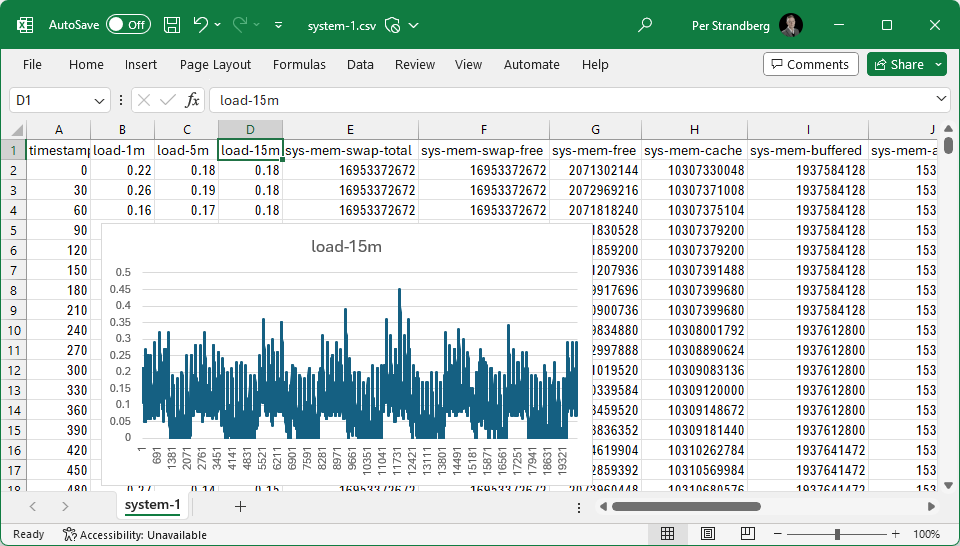}
    \caption{One CSV files loaded into MS Excel. The plot visualizes the first 20 thousand values for the load-15m metric. The plot represents about 0.05\% of the data set.}
    \label{fig:excel}
  \end{center}
\end{figure}

The data is stored in 19 CSV files, one per test system. These represent tables of data. Each file starts with a header of names of metrics. Each line then contains values. The first column contains the timestamp as seconds into the data collection. There are almost 86 thousand lines per file. Figure~\ref{fig:excel} illustrates the first rows and columns of one such file. In the remainder of this section we describe the meaning of the columns.

\subsubsection*{Timestamp}
\textit{A timestamp is a sequence of characters or encoded information identifying when a certain event occurred... They can have any epoch, can be relative to any arbitrary time, such as the power-on time of a system, or to some arbitrary time in the past.} --Wikipedia on Timestamp\footnote{\href{https://en.wikipedia.org/wiki/Timestamp}{https://en.wikipedia.org/wiki/Timestamp}}

Metric:
\begin{enumerate}
\item timestamp: Number of seconds since first data was collected.
\end{enumerate}

\subsubsection*{Load}
\textit{In UNIX computing, the system load is a measure of the amount of computational work that a computer system performs. The load average represents the average system load over a period of time.} --Wikipedia on System load\footnote{\href{https://en.wikipedia.org/wiki/Load_(computing)}{https://en.wikipedia.org/wiki/Load\_(computing)}}

Metrics:
\begin{enumerate}[resume]
\item load-1m: System load over the last 1 minute. 
\item load-5m: System load over the last 5 minutes.
\item load-15m: System load over the last 15 minutes.
\end{enumerate}

\subsubsection*{Memory}
\textit{Computer memory stores information, such as data and programs for immediate use in the computer... Besides storing opened programs, computer memory serves as disk cache and write buffer to improve both reading and writing performance. Operating systems borrow RAM capacity for caching so long as not needed by running software.} --Wikipedia on Computer memory\footnote{\href{https://en.wikipedia.org/wiki/Computer_memory}{https://en.wikipedia.org/wiki/Computer\_memory}}

\textit{In computer operating systems, memory paging (or swapping\ldots) is a memory management scheme by which a computer stores and retrieves data from secondary storage for use in main memory.} --Wikipedia on Memory paging\footnote{\href{https://en.wikipedia.org/wiki/Memory_paging}{https://en.wikipedia.org/wiki/Memory\_paging}}

Metrics (in bytes):
\begin{enumerate}[resume]
\item sys-mem-swap-total: Swap size (constant).
\item sys-mem-swap-free: Available swap.
\item sys-mem-free: Unused memory.
\item sys-mem-cache: Memory used for cache.
\item sys-mem-buffered: Memory used by kernel buffers.
\item sys-mem-available: Memory available to be allocated (free and cache).
\item sys-mem-total: Total size of memory (constant).
\end{enumerate}

\subsubsection*{CPU}
\textit{A central processing unit (CPU)... is the most important processor in a given computer. Its electronic circuitry executes instructions of a computer program, such as arithmetic, logic, controlling, and input/output (I/O) operations.}
-Wikipedia on Central processing unit\footnote{\href{https://en.wikipedia.org/wiki/Central_processing_unit}{https://en.wikipedia.org/wiki/Central\_processing\_unit}}

Metrics:
\begin{enumerate}[resume]
\item cpu-iowait: Summarized rate of change of seconds spent on waiting for I/O.
\item cpu-system: Summarized rate of change of seconds spent on kernel space threads.
\item cpu-user: Summarized rate of change of seconds spent on user space processes and threads.
\end{enumerate}

\subsubsection*{Disk/Storage}
These are metrics related to reading and writing (input and output) to and from a persistent storage (a disk). 

Metrics
\begin{enumerate}[resume]
\item disk-io-time: Rate of change in time spent on storage i/o operations.
\item disk-bytes-read: Rate of change in bytes read.
\item disk-bytes-written: Rate of change in bytes written.
\item disk-io-read: Rate of change in amount of read operations.
\item disk-io-write: Rate of change in amount of write operations.
\end{enumerate}

\subsubsection*{System}
\textit{In computing, ... fork is an operation whereby a process creates a copy of itself... Fork is the primary method of process creation on Unix-like operating systems.} 
--Wikipedia on fork (system call)\footnote{\href{https://en.wikipedia.org/wiki/Fork_(system_call)}{https://en.wikipedia.org/wiki/Fork\_(system\_call)}}

Metrics
\begin{enumerate}[resume]
\item sys-fork-rate: Rate of change in number of forks.
\item sys-interrupt-rate: Rate of change of interrupts.
\item sys-context-switch-rate: Rate of change of context switches.
\end{enumerate}

\subsubsection*{Thermal}
\textit{Temperature is a physical quantity that expresses quantitatively the attribute of hotness... The Celsius scale... is used... in most of the world...}--Wikipedia on Temperature\footnote{\href{https://en.wikipedia.org/wiki/Temperature}{https://en.wikipedia.org/wiki/Temperature}}

Metric
\begin{enumerate}[resume]
\item sys-thermal: Average rate of change in measured system temperature (Celsius). Please note that not all systems have access to a thermometer, so not all systems have this metric.
\end{enumerate}

\subsubsection*{Server up}
\textit{\ldots a heartbeat is a periodic signal generated by hardware or software to indicate normal operation or to synchronize other parts of a computer system\ldots Heartbeat messages are typically sent non-stop\ldots When the destination identifies a lack of heartbeat messages\ldots the destination may determine that the originator has failed\ldots}
--Wikipedia on Heartbeat (computing)\footnote{\href{https://en.wikipedia.org/wiki/Heartbeat_(computing)}{https://en.wikipedia.org/wiki/Heartbeat\_(computing)}}

Metric
\begin{enumerate}[resume]
\item server-up: freshness check of reporting server, values above 0 indicates that the server is available.
\end{enumerate}

\section{Experimental design, materials and methods}
The servers in the test systems run node exporter\footnote{\href{https://github.com/prometheus/node_exporter}{https://github.com/prometheus/node\_exporter}}, a tool for exporting performance data from servers or other PCs. Data was exported and stored using grafana\footnote{\href{https://grafana.com/}{https://grafana.com/}}. With a Python script, data was exported and slightly cleaned into CSV files.

The servers are located in the software testing laboratory of Westermo Network Technologies AB, in Västerås, Sweden.

\section{Ethics statements}
To protect Westermo, data has been anonymized, obfuscated, and only a sub-set of available test systems has been used. The public release of data has been approved by staff responsible for information security at Westermo Network Technologies AB at risk workshops during 2023.

\section{CRediT author statement}
The authors of this paper has contributed as follows.
Conceptualization: PES,
Methodology: PES \& YM,
Software: YM,
Validation: PES \& YM,
Investigation: PES \& YM,
Data Curation: YM,
Writing - Original Draft: PES,
Writing - Review \& Editing: PES \& YM,
Visualization: PES,
and
Supervision: PES.
Other Westermo staff has contributed with Resources and Software.

\section{Acknowledgments}
Work was funded by 
Westermo Network Technologies AB,
and the AIDOaRt project, a European ECSEL Joint Undertaking (JU) under grant agreement No.\ 101007350.

\section{Declaration of interests}
The authors of this paper are employed at Westermo Network Technologies AB.

\section{License}
This data set is licensed with the 
\href{https://creativecommons.org/licenses/by/4.0/}{Creative Commons Attribution 4.0 International}.

In short, you are free to: share, copy and redistribute the material; and to adapt, remix, transform, and build upon it for any purpose;
under the condition that you you give appropriate credit, and do not restrict others from doing anything the license permits. Read the license for details.

Suggested attribution:
P E Strandberg and Y Marklund. (2023). The Westermo test system performance data set. Retrieved from \href{https://github.com/westermo/test-system-performance-dataset}{https://github.com/westermo/test-system-performance-dataset}

\bibliographystyle{abbrv}
\bibliography{references.bib} 

\end{document}